\documentclass{JINST}

\title{Automatic track recognition for large-angle minimum ionizing particles in nuclear emulsions}

\author{T. Fukuda$^a$\thanks{Corresponding
author.}~, S. Fukunaga$^a$, H. Ishida$^a$, T. Matsumoto$^a$, T. Matsuo$^a$, S. Mikado$^b$,\newline S. Nishimura$^a$, S. Ogawa$^a$, H. Shibuya$^a$, J. Sudou$^a$, A. Ariga$^c$ and S. Tufanli$^c$\\
\llap{$^a$}Fundamental Physics Laboratory, Toho University,\\
  Miyama, Funabashi J-274-8510, Japan\\
\llap{$^b$}Nihon University,\\
  Narashino J-275-8576, Japan\\
\llap{$^c$}Laboratory for High Energy Physics (LHEP), University of Bern,\\
  Bern CH-3012, Switzerland\\
  E-mail: \email{tsutomu.fukuda@ph.sci.toho-u.ac.jp}}

\abstract{We previously developed an automatic track scanning system which enables the detection of large-angle nuclear fragments in the nuclear emulsion films of the OPERA experiment. As a next step, we have investigated this system's track recognition capability for large-angle minimum ionizing particles $(1.0 \leq |tan \theta| \leq 3.5)$. This paper shows that, for such tracks, the system has a detection efficiency of 95$\%$ or higher and reports the achieved angular accuracy of the automatically recognized tracks. This technology is of general purpose and will likely contribute not only to various analyses in the OPERA experiment, but also to future experiments, e.g. on low-energy neutrino and hadron interactions, or to future research on cosmic rays using nuclear emulsions carried by balloons.}


\begin{document}

\section{Introduction}

\begin{figure}[t]
\begin{center}
\includegraphics[clip, width=11.0cm]{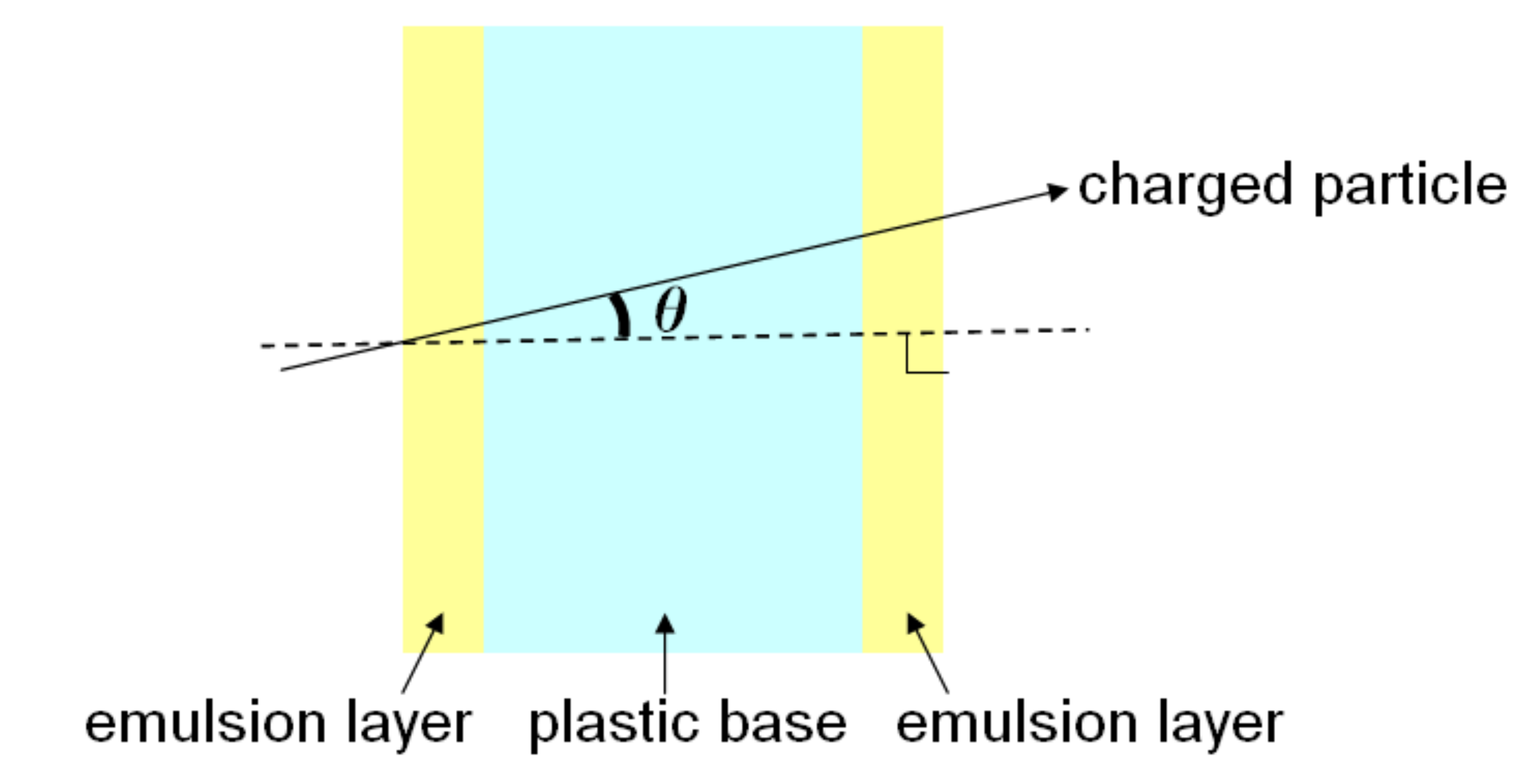}
\caption{Cross-section of a typical nuclear emulsion film. The angle of a charged particle track recorded in the emulsion layers is indicated as tan$\theta$ , with $\theta$ being the angle to the axis perpendicular to the film.}
\label{fig-1}
\end{center}
\end{figure}

        Nuclear emulsions are 3-dimensional solid tracking detectors with sub-micrometric spatial resolution. Gelatin layers with a thickness of a few 10 $\mu$m to 1 mm are coated onto both sides of a plastic base and AgBr crystals with diameter of approx. 200 nm are incorporated into the material as photosensitive detectors (figure \ref{fig-1}). When charged particles pass through the emulsion, latent image specks are formed due to the ionization and are amplified in the development process as silver grains of diameter approx. 0.6 $\mu$m. If the detector is then observed with an optical microscope, it is possible to detect the tracks of the charged particles as lines of silver grains. Nuclear emulsion has sensitivity for the charged particles emitted in all directions and is suitable for precise and thorough investigation of the primary interaction points in elementary particle collisions. Many physics results have been achieved using these nuclear emulsions (\cite{bib1}-\cite{bib13}).

        In modern analysis of nuclear emulsions, the analysis speed has been dramatically improved due to the development of devices to automatically scan nuclear emulsions, referred to here as "track selectors" or TS (\cite{bib14}-\cite{bib18}). In recent experiments with nuclear emulsions, the research was focused on the detection of particles produced in high-energy collisions. The emission of such particles is concentrated at small angles and thus almost no studies have been done on automatic recognition of large-angle particle tracks. In order to analyze the OPERA experiment, we have developed new TS enabling the recognition of large-angle tracks \cite{bib19} and conducted a systematic analysis of the production of large-angle nuclear fragments from hadron interactions \cite{bib20}. This system enables large-angle track detection using a wide-field objective lens. In addition we solved the problem of the much larger amount of track recognition processing induced by the widening of the angular range. This was achieved for the first time, by massive parallel processing using a Graphics Processing Unit or GPU, and thereby increasing the processing speed. While at first we had designed the system to measure only nuclear fragments with high ionization loss, but in the course of this study, it was discovered that the recognition of tracks works at high efficiency even for large-angle minimum ionizing particles (MIP). Thus a dedicated beam test experiment was carried out at CERN to systematically investigate automatic track recognition of large-angle MIPs within up to $|tan \theta|$ = 3.5 (about 5 times larger than the conventional solid angular acceptance). We measured the track recognition efficiency and the angular accuracy of new system in dependence of the track angle. The results will be presented in this paper.

\section{Evaluation method of the large-angle tracking}

\subsection{Detector and beam exposure}

\begin{figure}[t]
\begin{center}
\includegraphics[clip, width=7.0cm]{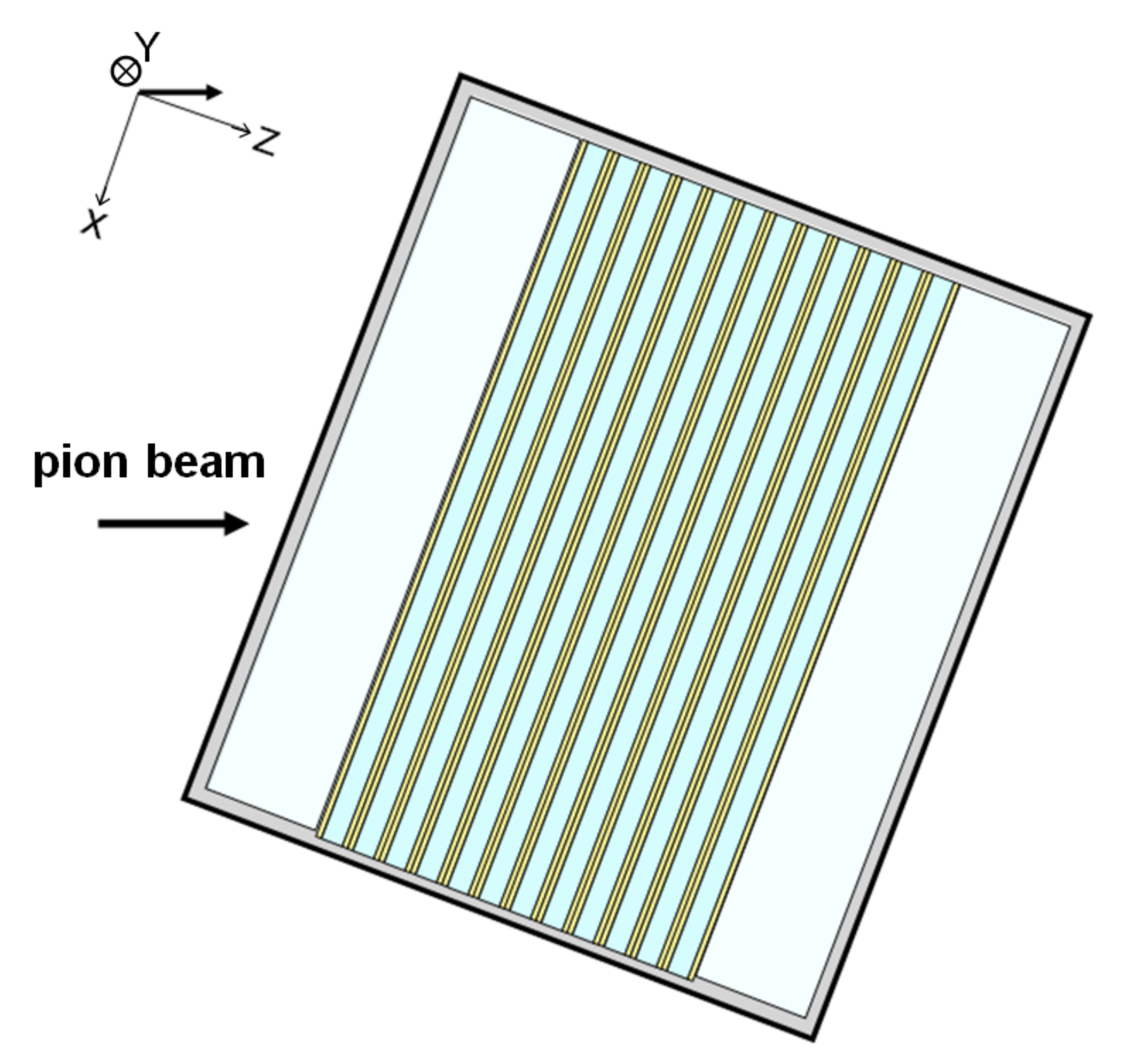}
\caption{Beam exposure of the detector. At each exposure step, the inclination of the detector relative to the beam is changed.}
\label{fig-2}
\end{center}
\end{figure}

        The detector modules used in this experiment consisted of the OPERA nuclear emulsion films \cite{bib21}, where 205 $\mu$m-thick plastic bases are coated on both sides with emulsion layers of 44 $\mu$m thickness. The OPERA films have outstanding uniformity in terms of the emulsion layer thickness and size of the contained AgBr crystals. The grain density which indicates the sensitivity of the emulsion is $\sim$33 grains/100 $\mu$m. The film size is 12.5 $\times$ 10.0 $cm^2$. As shown in figure \ref{fig-2}, 12 sheets of OPERA film were sandwiched from both sides by 1 mm acrylic plates to ensure planarity and were stabilized by placing them under vacuum in an aluminium pack. The total detector thickness was about 0.55 cm, representing about 1 percent of a hadronic interaction length.

        The beam exposure was carried out on the CERN T10 beam line in August 2012. The detector was exposed to the 6 GeV/c negative charged pion beam ($\pi^-$) with 24 different incident angles. As shown in figure \ref{fig-2}, tracks at different angles were recorded by rotating the detector at each exposure step. Altogether 24 different incident track angles were realized, all at $tan \theta_y \sim 0$ and at $tan \theta_x$ varying from -2.2 to 3.5. The density of beam tracks recorded at each exposure step is between $10^2 /cm^2$ and $10^3 /cm^2$. Then all measured beam tracks are clearly distinguishable by position and angle information of each track.

\subsection{Scanning and data analysis}

        The main steps of the track recognition algorithm implemented in our TS are ( see section 2 of \cite{bib19}  for details ):

\begin{enumerate}
\item The 44 $\mu$m emulsion layer is captured as a 16 tomographic images.
\item For each image its own smoothing image is subtracted to correct the non-uniformity of brightness.
\item The multilevel image is binarized by setting a brightness threshold for each pixel.
\item If there is a pixel above the threshold which is defined as "hit", hits at its surrounding pixels are created (expansion processing: typically 1 hit is expanded to 3 $\times$ 3 pixels ; 1 pixel is set at 0.275 $\mu$m ).
\item For each hypothesis on the track angles ($\theta_x, \theta_y$) in the setup angular acceptance, the 16 processed images are searched for a series of aligned grains created by a penetrated charged particle, i.e. track elements.
\item The total number of images where a hit associated to a track is present is called the pulse height (PH) of the track elements. The valid track elements are selected by applying a PH threshold of 6 or 7.
\end{enumerate}

\begin{figure}[t]
\begin{center}
\includegraphics[clip, width=15.0cm]{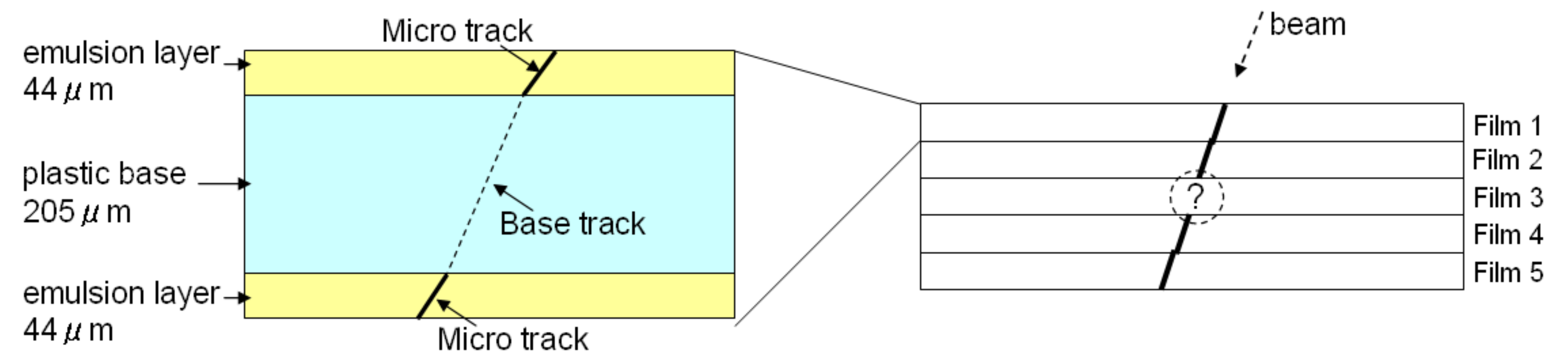}
\caption{Schematic view of the track reconstruction. The diagram at left shows a cross-section of one emulsion film. The diagram at right is a cross-section of 5 consecutive films. The requirement of having at least 4 base tracks connected over in 5 films eliminates the background due to low-energy particles, such as electrons from environmental radiation.}
\label{fig-3}
\end{center}
\end{figure}

\begin{figure}[hb]
\begin{center}
\includegraphics[clip, width=10.0cm]{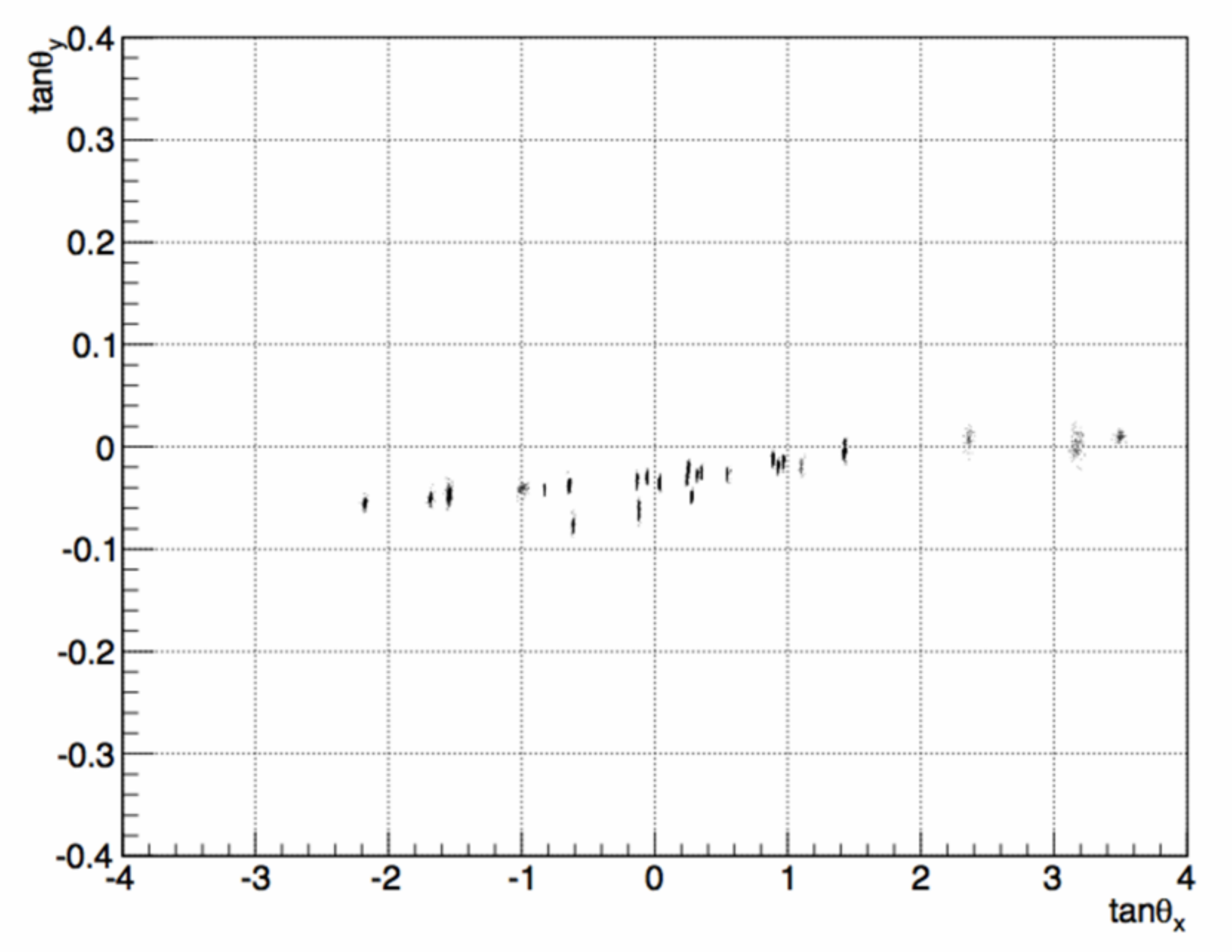}
\caption{Angular distribution of the track samples.}
\label{fig-4}
\end{center}
\end{figure}

        The scanning was done on 1 $cm^2$ of both sides of the 5 upstream films with the conditions |$tan \theta| <$ 4.0 and PH = 7. As shown on the left in figure \ref{fig-3}, a track element in one emulsion layer is called a "micro track" and the reconstructed path through the plastic base which connects positions of micro tracks on both sides of the film is called a "base track". A micro track is reconstructed by using the hits of the first and the last images. The hit position in the nearest image from the plastic base is used as the position of the micro track. Then the angle of base track is calculated by using the micro track positions at both surfaces of the plastic base. In order to select MIP tracks passing through the exposed films, the base tracks of all films were connected (\cite{bib22}, \cite{bib23}), and tracks with base tracks in films 1, 2, 4, and 5 were selected in the analysis sample (right side of figure \ref{fig-3}). The track detection efficiency per film was then found by checking the presence of a base track in film 3. In this analysis the evaluation was conducted for each beam angle spot by randomly selecting 100 - 200 beam tracks. Figure \ref{fig-4} shows the angular distribution of the selected track samples.

\section{Performance of the large-angle track recognition}

\subsection{Angular accuracy of micro tracks}

\begin{figure}[htp]
\begin{center}
\includegraphics[clip, width=14.0cm]{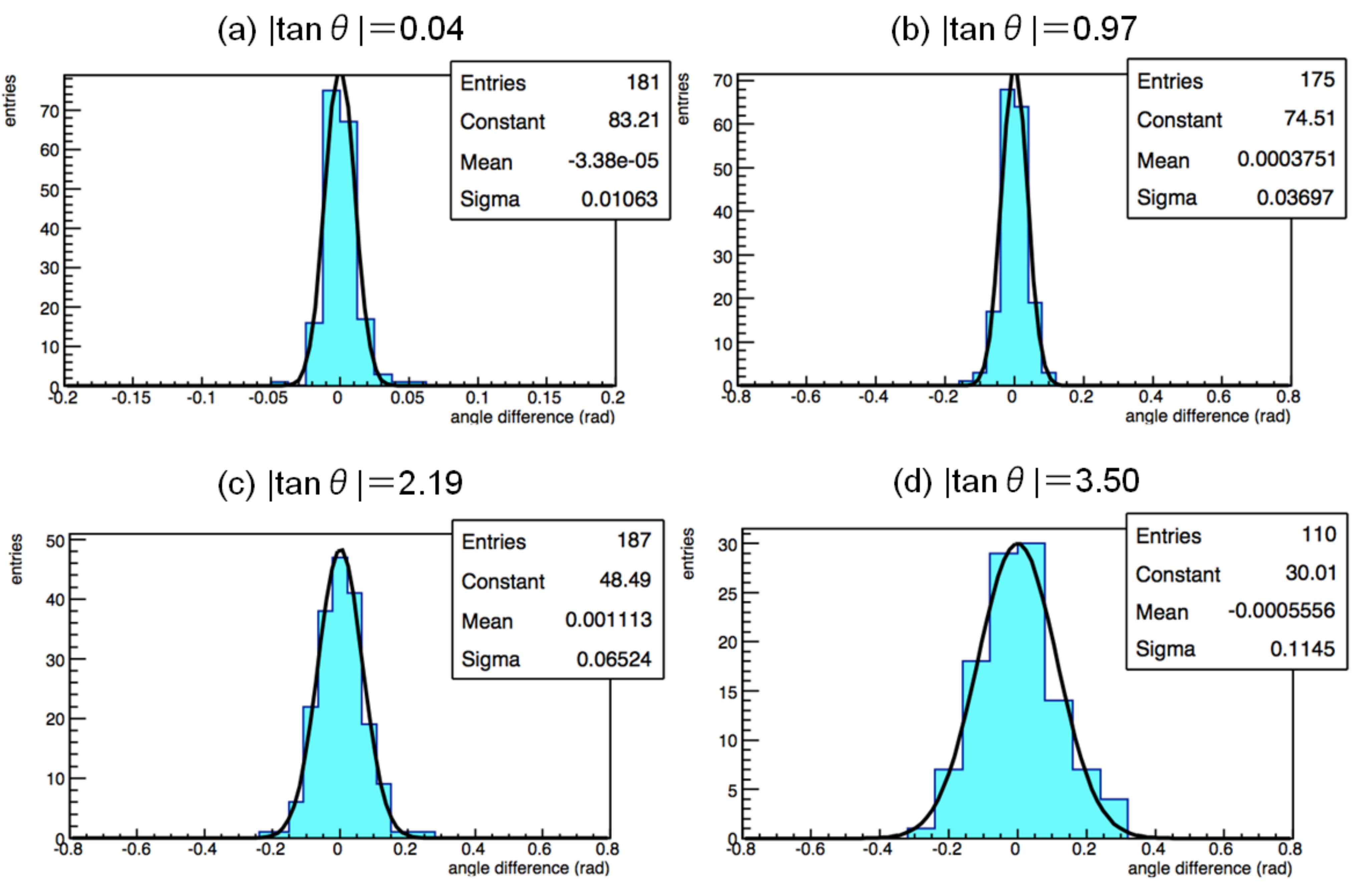}
\caption{Angle difference between micro tracks and its corresponding base track. (a) for |$tan \theta$| = 0.04, (b) for |$tan \theta$| = 0.97, (c) for |$tan \theta$| = 2.19, (d) for |$tan \theta$| = 3.50.}
\label{fig-5}
\end{center}
\end{figure}

        The angular accuracy of the recognized micro tracks is indicated by the spread of the distribution of the angle difference between base track and micro track, as shown in figure \ref{fig-5}. For |$tan \theta$| = 0.04 the accuracy (1$\sigma$) is 10.6 mrad, but as the track angle increases, e.g. for |$tan \theta$| = 0.97, 2.19, and 3.50, the accuracy becomes worse: 37.0 mrad, 65.2 mrad, and 114.5 mrad, respectively. Figure \ref{fig-6} shows the angular accuracy (1$\sigma$) at each measured angle. An approximately linear dependence is observed. 

\subsection{Angular accuracy of base tracks}

\begin{figure}[htp]
\begin{center}
\includegraphics[clip, width=14.0cm]{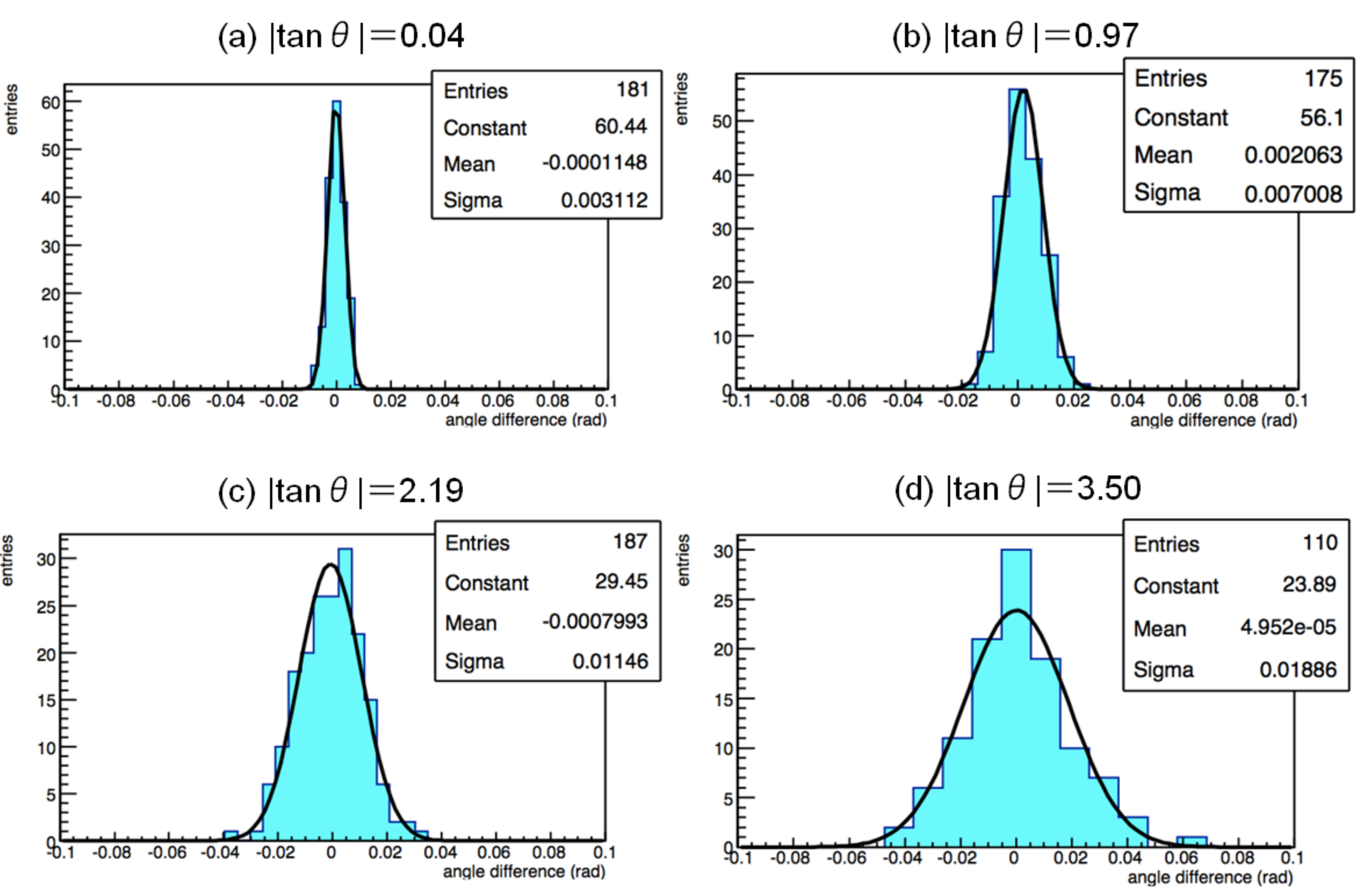}
\caption{Angle difference of base tracks in film 3 from their corresponding high-energy tracks penetrating 5 emulsion films. (a) for |$tan \theta$| = 0.04, (b) for |$tan \theta$| = 0.97, (c) for |$tan \theta$| = 2.19, (d) for |$tan \theta$| =3.50.}
\label{fig-7}
\end{center}
\end{figure}

        The angular accuracy of the reconstructed base tracks is calculated as the spread of the distribution of the angle differences between the base track in film 3 and the straight line obtained by fitting the positions of the base tracks in films 1, 2, 4, and 5. As can be seen in figure \ref{fig-7}, the angular accuracy of base tracks behaves like the angular accuracy of micro tracks: The angular accuracy (1$\sigma$) is 3.1 mrad for |$tan \theta$| = 0.04 and, as the track angle increases, e.g. for |$tan \theta$| = 0.97, 2.19, and 3.50, it gets larger: 7.0 mrad, 11.5 mrad, and 18.9 mrad, respectively. The angular accuracy of base tracks is smaller than that of micro tracks because of the long distance between connected points. Figure \ref{fig-8} shows the angular accuracy (1$\sigma$) at each measured angle. Here also a linear dependence is observed. These dependence will further be discussed in section \ref{sec:4.1}.

\subsection{Track recognition efficiency}

        Figure \ref{fig-9} shows the PH distribution of micro tracks attached to a reconstructed base track while figure \ref{fig-10} shows the angle dependence of the base track detection efficiency. It amounts to 90$\%$ around |$tan \theta$| = 1.0 and rises as the angle increases. This means that the micro track detection efficiency is more than 95$\%$ for large-angle tracks. The interpretation of this angle dependence of the detection efficiency will further be discussed in section \ref{sec:4.2}.

\section{Discussions}

\subsection{Discussion of the angle dependence of the angular accuracy} \label{sec:4.1}

\begin{figure}[hbp]
\begin{center}
\includegraphics[clip, width=11.0cm]{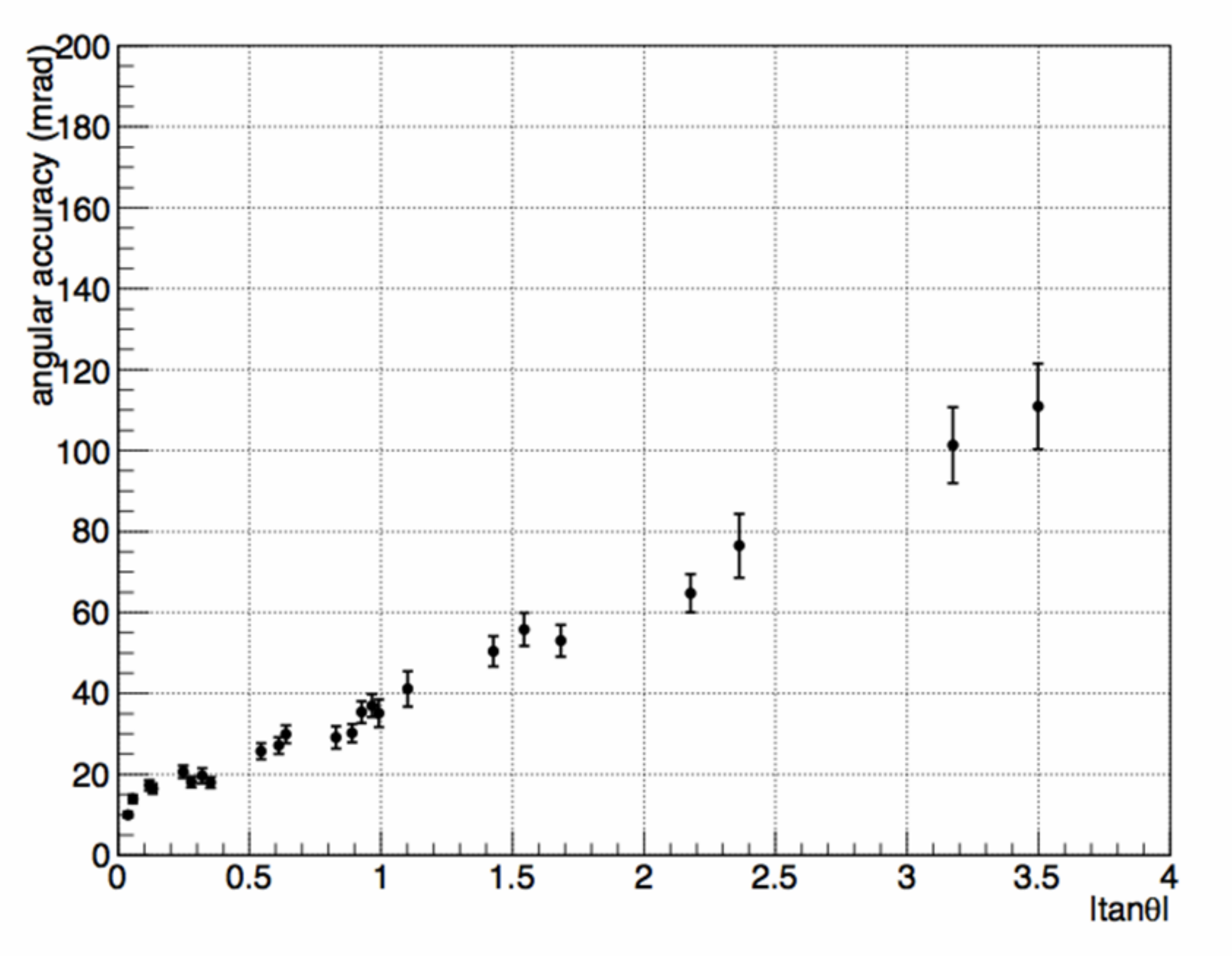}
\caption{Angle dependence of the angular accuracy of micro tracks.}
\label{fig-6}
\end{center}
\end{figure}

\begin{figure}[hbp]
\begin{center}
\includegraphics[clip, width=11.0cm]{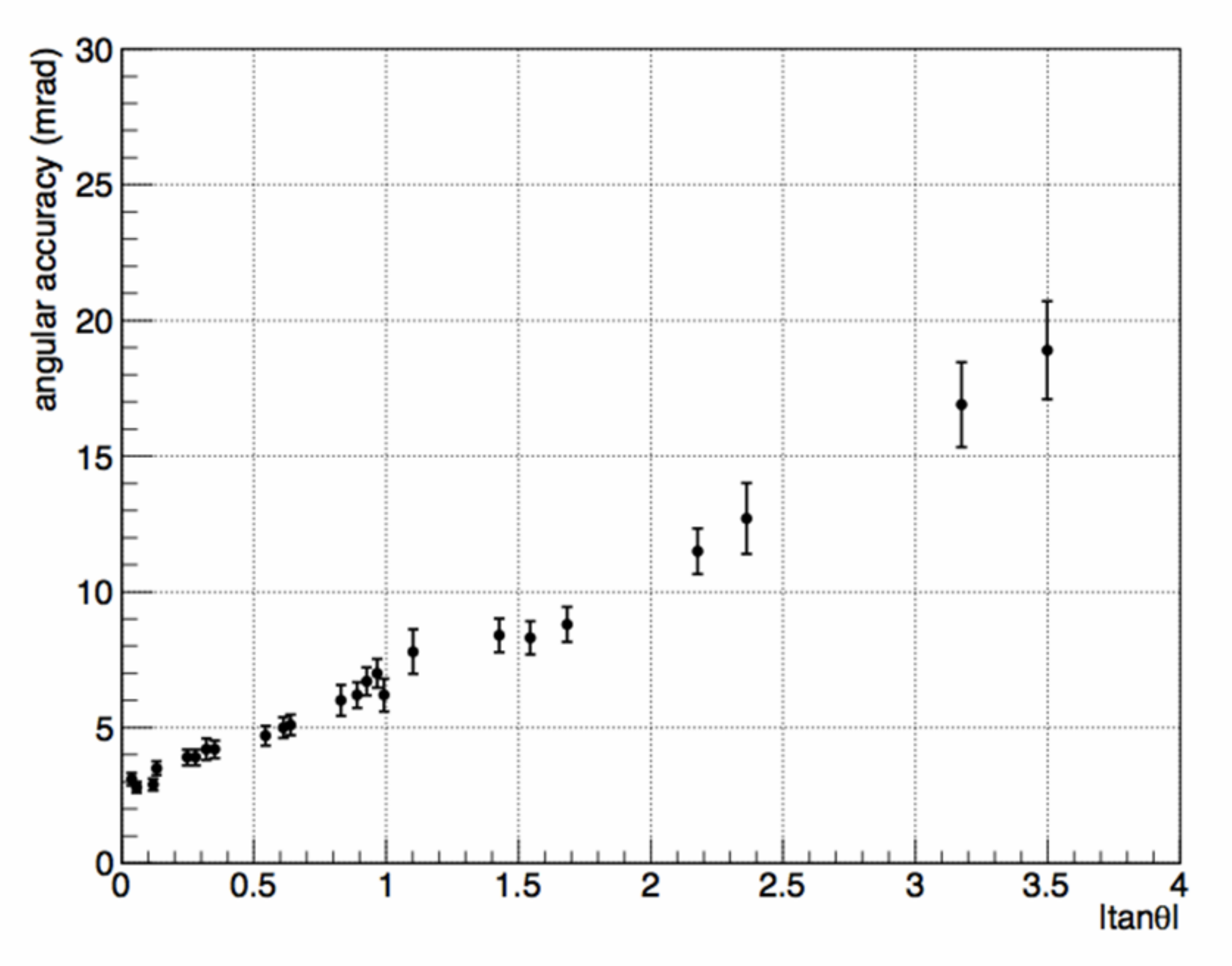}
\caption{Angle dependence of angular accuracy of base tracks.}
\label{fig-8}
\end{center}
\end{figure}

        The slope $tan \theta_x$ of a track is defined by measuring the start position ($x_1, z_1$) and end position ($x_2, z_2$) of the track (eq. \ref{eq-1}). $\Delta z$ of a micro track and that of a base track are 44 $\mu$m and 205 $\mu$m, respectively. Therefore the angular accuracy can be expressed by eq. \ref{eq-2} depending on the position measurement accuracy, $\delta x$, and the measurement accuracy of the emulsion layer and plastic base surface, $\delta z$. Fitting the results shown in figure \ref{fig-6} and \ref{fig-8} with eq. \ref{eq-2} gives $\delta x \sim$0.40 $\mu$m and $\delta z \sim$1.0 $\mu$m for micro tracks and $\delta x \sim$0.46 $\mu$m and $\delta z \sim$0.8 $\mu$m for base tracks. The values calculated for micro tracks and base tracks are similar and also consistent with the specifications of our system.

\begin{eqnarray}
tan \theta&=&\frac{x_2-x_1}{z_2-z_1}=\frac{\Delta x}{\Delta z} \label{eq-1}\\
\sigma^2_{tan \theta}&=&\frac{2}{\Delta z^2} (\delta x^2+\delta z^2 \times tan^2 \theta) \label{eq-2}
\end{eqnarray}

\bigskip

\bigskip

\begin{figure}[htp]
\begin{center}
\includegraphics[clip, width=14.0cm]{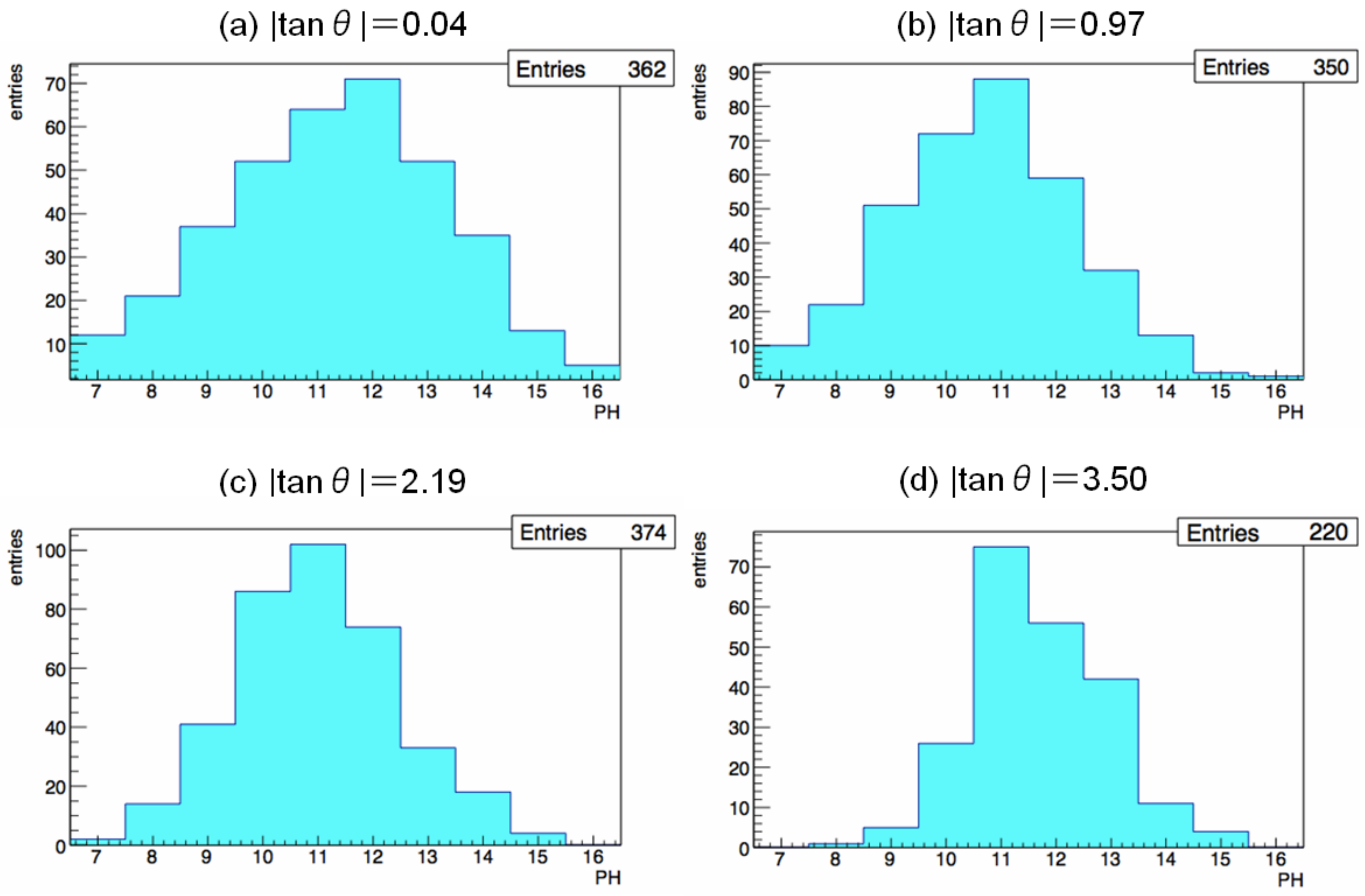}
\caption{PH distribution of micro tracks. (a) for |$tan \theta$| = 0.04, (b) for |$tan \theta$| = 0.97, (c) for |$tan \theta$| = 2.19, (d) for |$tan \theta$| = 3.50.}
\label{fig-9}
\end{center}
\end{figure}

\begin{figure}[hbp]
\begin{center}
\includegraphics[clip, width=11.5cm]{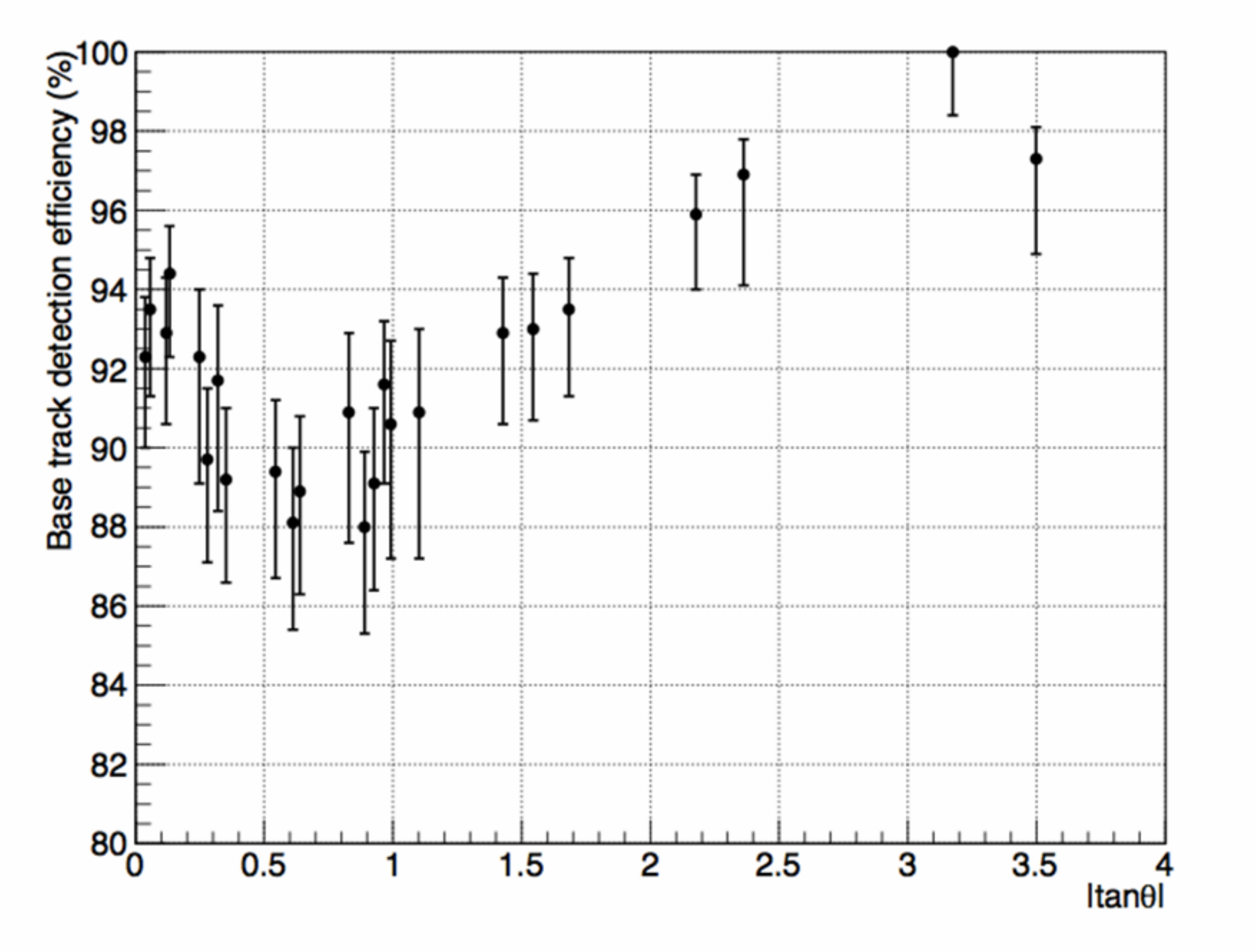}
\caption{Angle dependence of the base track detection efficiency.}
\label{fig-10}
\end{center}
\end{figure}

\subsection{Discussion of the angle dependence of the track recognition efficiency} \label{sec:4.2}

        In the conventional angle measurement range (|$tan \theta$| $\le$ 0.6) it has been observed that, in automatic track recognition under the condition PH $\ge$ 7, the detection efficiency decreases as the angle increases \cite{bib18}. This can be explained as follows: the z coordinate of a hit recorded in an emulsion layer is taken as the z-value of the focal plane of this layer. It can deviate from the true z coordinate of the silver grain by as much as half the microscope's focal depth, i.e. 2-3 $\mu$m. For tracks with a non-negligible theta angle this deviation may yield to the rejection of the hit as not aligned with the hits in the other layers. The PH distribution is thus shifted to lower values and, consequently, the track detection efficiency decreases. The same tendency is also observed in the present analysis, as shown in figure \ref{fig-9} (a), (b) and \ref{fig-10}, but, in addition, it was also found that the detection efficiency begins to rise when the measurement range is extended beyond the conventional angle measurement range. The interpretation of this effect is that, at very large angles, the charged particle passes through a longer emulsion distance and thus produces a larger number of silver grains per layer. The mean value of the number of silver grains recorded when passing an emulsion layer of 42 $\mu$m thickness perpendicularly is 14. This value becomes 20 when |$tan \theta$| = 1.0, 31 when |$tan \theta$| = 2.0, and 44 when |$tan \theta$| = 3.0. Since the track length in an emulsion layer varies as $\sqrt{(1+tan^2 \theta)}$, this effect starts to play a dominant role from about |$tan \theta$| = 1.0. This is consistent with the trend that can be seen in figure \ref{fig-10}. Let us note however a possible decrease of the efficiency at |$tan \theta$| = 3.5. Our next study of tracks with even larger angles will help to clarify this point.

\section{Applications}

        This technology described above has already been applied in the framework of the OPERA experiment to improve the reliability of tau neutrino candidate selection (\cite{bib11}-\cite{bib13}, \cite{bib20}).

        Also a small number of interactions with atmospheric neutrinos coming from the southern hemisphere may be identified in the OPERA detector. Since the films are perpendicular to the neutrino beam from CERN, such cosmic ray events will be approximately parallel to the emulsion film. Therefore the fact that large-angle MIP tracks can be automatically measured with high efficiency is an important point in discussing the feasibility of this kind of research, for example clear tau neutrino appearance event search in the atmospheric neutrino oscillation. Furthermore this technology is also assumed to be very useful in the search for exotic events caused by cosmic rays.

        In future studies on low-energy neutrino/hadron interactions in the sub-GeV range also using nuclear emulsions, the emitted particles will have low energy, and thus the track angles will be large. Therefore automatic recognition of large-angle tracks will be a powerful tool. Also in cosmic ray experiments where nuclear emulsions are loaded onto balloons \cite{bib24}, a wider angle measurement range will give much larger acceptance for event selection. As another example of applications, when conducting a re-analysis of the JACEE experiment \cite{bib25}, this new technique will enable a much more efficient analysis by directly detecting the large-angle particles emitted from the pair annihilation points of anti-protons and anti-matter.

\section{Conclusions}

        This paper summarized the results of the improved automatic track measurement of large-angle minimum ionizing particles in nuclear emulsions. In systematically analyzing minimum ionizing particles within up to 5 times the conventional solid angular acceptance, the automatic measurement resulted in efficiencies of 95 $\%$ or higher. This paper has also summarized the results on the angle dependence of the track detection efficiency and the angular accuracy. The technology we developed for the automatic detection of large-angle tracks is of general purpose, and we briefly presented its possible application in various future nuclear emulsion analyses.

\acknowledgments

        We appreciate the support provided by the Toho University and the collaborating laboratories in the OPERA experiment. We acknowledge the support from the Japan Society for the Promotion of Science (JSPS) through their grants (JSPS KAKENHI Grant Number 23740184, 25707019). We also gratefully acknowledge P. Vilain and A. Hollnagel for their careful reading of the manuscript.

\end{document}